\documentclass[11pt]{article}
\usepackage{amssymb}
\usepackage{amsmath}
\usepackage{amscd}
\usepackage{latexsym}

\catcode`@=11
\renewcommand{\section}{\@startsection{section}{1}{0pt}{\medskipamount}
{\medskipamount}{\large\bf}}
\numberwithin{equation}{section}
\catcode`@=12
\def\a{\alpha}
\def\b{\beta}
\def\g{\gamma}

\def\de{\delta}

\def\th{\theta}

\def\la{\lambda}
\def\m{\mu}
\def\n{\nu}
\def\r{\rho}
\def\s{\sigma}
\def\p{\phi}
\def\vp{\varphi}

\def\1{\bar 1}
\def\2{\bar 2}
\def\3{\bar 3}
\def\Ups{\Upsilon}
\def\eps{\epsilon}

\newcommand{\yb}{\bar{y}}

\newcommand{\C}{\mathbb C}
\newcommand{\R}{\mathbb R}
\newcommand{\T}{\mathbb T}
\newcommand{\Z}{\mathbb Z}

\newcommand{\Acal}{{\cal A}}
\newcommand{\Ncal}{{\cal N}}
\newcommand{\Lcal}{{\cal L}}
\newcommand{\Fcal}{{\cal F}}
\newcommand{\Ecal}{{\cal E}}

\def\im{\mbox{i}}
\def\N2{$N{=}2$}
\def\pa{\mbox{$\partial$}}
\def\diff{\mbox{d}}
\def\tr{{\rm tr}}
\def\sfrac#1#2{{\textstyle\frac{#1}{#2}}}
\def\>{\rangle}
\def\<{\langle}
\def\+{\dagger}
\def\={\ =\ }

\def\sn{\mbox{sn}}

\begin{document}
\begin{titlepage}
\setcounter{page}{0}
\begin{flushright}
March 2008\\
ITP-UH-07/08
\end{flushright}

\vskip 1.2cm

\begin{center}

{\Large\bf
Explicit Non-Abelian Monopoles and Instantons\\[6pt]
in SU($N$) Pure Yang-Mills Theory}

\vspace{10mm}
{\large Alexander D. Popov$^*$}
\\[3mm]
\noindent {\em Institut f\"ur Theoretische Physik,
Leibniz Universit\"at Hannover \\
Appelstra\ss{}e 2, 30167 Hannover, Germany }
\\
{Email: {\tt popov@itp.uni-hannover.de}}
\\[2mm]
and
\\[2mm]
\noindent {\em Bogoliubov Laboratory of Theoretical Physics, JINR\\
141980 Dubna, Moscow Region, Russia}
\\
{Email: {\tt popov@theor.jinr.ru}}
\vspace{10mm}

\begin{abstract}
\noindent
It is well known that there are no static non-Abelian monopole solutions in pure Yang-Mills
theory on Minkowski space $\R^{3,1}$. We show that such solutions exist in SU($N$) gauge
theory on the spaces $\R^2\times S^2$ and $\R\times S^1\times S^2$ with Minkowski signature
$(-+++)$. In the temporal gauge they are solutions of pure Yang-Mills theory on $\T\times S^2$,
where $\T$ is $\R$ or $S^1$. Namely, imposing SO(3)-invariance and some reality conditions, we
consistently reduce the Yang-Mills model on the above spaces  to a non-Abelian analog of
the $\p^4$ kink model whose static solutions give SU($N$) monopole (-antimonopole)
configurations on the space $\R^{1,1}\times S^2$ via the above-mentioned correspondence.
These solutions can also be considered as instanton configurations of Yang-Mills theory
in $2+1$ dimensions.
The kink model on $\R\times S^1$ admits also periodic sphaleron-type solutions describing
chains of $n$ kink-antikink pairs spaced around the circle $S^1$ with arbitrary $n>0$.
They correspond to chains of $n$ static monopole-antimonopole pairs on the space
$\R\times S^1\times S^2$ which can also be interpreted as instanton configurations
in 2+1 dimensional pure Yang-Mills theory at finite temperature (thermal time circle).
We also describe similar solutions in Euclidean SU($N$) gauge theory on $S^1\times S^3$
interpreted as chains of $n$ instanton-antiinstanton pairs.
\end{abstract}
\end{center}

\vfill

\textwidth 6.5truein
\hrule width 5.cm

{\small
\noindent ${}^*$
Supported in part by the Deutsche Forschungsgemeinschaft (DFG).}

\end{titlepage}

\section{Introduction and summary}

\noindent
Magnetic monopoles~\cite{1} are playing an important role in the nonperturbative
physics in 3+1 dimensional Yang-Mills-Higgs theory~\cite{2}-\cite{5}. In particular,
quark confinement is believed to be explained by the condensation of monopoles
and the dual Meissner effect~\cite{6}. Confinement should also be a property of the
pure Yang-Mills theory without any matter. However, there are no non-Abelian monopoles
in pure gauge theory on $\R^3$ and that is why the dual Meissner effect is discussed
in terms of the Dirac monopoles (`Abelian dominance'), whose embedding into SU(2)
Yang-Mills are point (singular) Wu-Yang monopoles~\cite{7}.\footnote{See~\cite{8} for
a multi-monopole generalization of the SU(2) Wu-Yang solution with monopoles located
at arbitrary points in $\R^3$.}

In this paper, we construct explicit smooth static monopole and monopole-antimonopole
solutions in SU($N$) pure Yang-Mills theory on the space $\R\times\T\times S^2$
with Minkowski signature $(-+++)$, where $\T$ is $\R$ or $S^1$. These configurations
can be considered as finite action solutions in pure Yang-Mills theory on the
Euclidean space $\T\times S^2$ (temporal gauge), i.e. as instanton configurations.
We also describe explicit instantons
and chains of instanton-antiinstanton pairs on the Euclidean spaces $\R\times S^3$
and $S^1\times S^3$, respectively. Note that instanton-antiinstanton chains on $\R^4$
describe certain multiparticle scattering events leading to the transitions between
topologically distinct vacua~\cite{9}.\footnote{For further discussion of the role of
instantons in field theory and references see e.g.~\cite{10}.}  On the other hand,
it was recently shown that SU($N$) gauge theories on $S^1\times S^{d-1}$
demonstrate for large $N$ and weak coupling a confinement-deconfinement transition
at temperatures proportional to the inverse scale of the spheres~\cite{11,12}. All
this may justify the study of the above-mentioned solutions to the Yang-Mills equations
on the spaces $\T\times S^{d-1}$ with $d=3,4$.

The outline of this paper is as follows. We start our discussion in Section 2 with
SU(2) Yang-Mills theory on the space $\Sigma^{1,1}\times S^2$ with Minkowski signature
$(-+++)$. Here $\Sigma^{1,1}$ is the space $\R^{1,1}$ or $\R\times S^1$ and
$S^2=\,$SO(3)/SO(2) is the standard two-sphere. We consider SO(3)-invariant gauge
field with the SO(3)-invariance defined up to a gauge transformation (cf.~\cite{13}).
The imposition of this symmetry condition reduces SU(2) Yang-Mills theory on
$\Sigma^{1,1}\times S^2$ to the Abelian Higgs model in 1+1 dimensions and, in the case
of vanishing gauge field on $\Sigma^{1,1}$, to the standard $\p^4$ kink model on the
space $\Sigma^{1,1}$, which are discussed in Section 3. Static kink and antikink
configurations on $\R^{1,1}$ give static monopole and antimonopole
solutions\footnote{However, Lorentz rotations in  $\R^{1,1}$ yield moving
dyons from the viewpoint of 3+1 dimensions.} of the SU(2) Yang-Mills equations on
$\R^{1,1}\times S^2$, which can also be interpreted as instanton and antiinstanton
configurations in Yang-Mills theory in $2+1$ dimensions.  In the $\p^4$ kink model on
$\R\times S^1$ there exist static sphaleron configurations (chains of alternating kinks
and antikinks equally spaced around the circle $S^1$~\cite{14}) corresponding in our
construction to chains of monopole-antimonopole pairs on the space $S^1\times S^2$.
The above-mentioned relation between kinks and sphalerons in 1+1 dimensions and SU(2)
monopoles and chains of monopole-antimonopole pairs in 3+1 dimensions is discussed in
Section 4. In Section 5 we generalize our results  to the case of SU($N$) pure Yang-Mills
theory on the space $\Sigma^{1,1}\times S^2$. Namely, for any $N>2$ the Yang-Mills equations
are reduced to a non-Abelian analog of $\p^4$ kink equation which can be interpreted as
describing $N-1$ interacting kinks and antikinks. It has multi-kink and kink-antikink
static solutions. These solutions correspond to SU($N$) multi-monopole and
monopole-antimonopole configurations on the spaces $\R\times S^2$ and  $S^1\times S^2$.
Finally, in Section 6 we consider SU($N$) Yang-Mills equations on the Euclidean space
$\T\times S^3$ and their SO(4)-symmetric reduction to a matrix model on $\T$, where $\T$
is $\R$ or $S^1$. We discuss further algebraic reduction of the matrix field equations
to Toda-like equations and describe some of their solutions. They are instantons on the
space $\R\times S^3$ and chains of $n$ instantons and $n$ antiinstantons on the space
$S^1\times S^3$. Possible applications of these solutions and their feasible generalizations
are briefly discussed.

\section{Yang-Mills theory on $\Sigma^{1,1}\times S^2$}

\noindent
{\bf Manifold $\Sigma^{1,1}\times S^2$.} Let  $\Sigma^{1,1}$ be the
space $\R^2$ or $\R\times S^1$ with (local)  real coordinates $x^i$ and
Minkowski metric $\eta =(\eta_{ij})= \mbox{diag}(-1,+1)$, where $i,j,...=0,3$.
We use this unusual numbering since it is more convenient from the
viewpoint of $3+1$ dimensions where axial symmetry is usually related
with the $x^3$-axis. Consider the product $\Sigma^{1,1}\times S^2$
with local real coordinates $x^\m$, where indices $\m ,\n ,...$ run
through 0,1,2,3 so that $x^1$ and $x^2$ are local coordinates on the
two-sphere $S^2$.

We also introduce on $S^2\cong\C P^1$ a local complex coordinate
$y=\sfrac12(x^1+\im x^2)$ and coordinates $0\le\th <\pi$,
$0\le\vp\le 2\pi$ via
\begin{equation}\label{2.1}
y=R\tan\Bigl(\frac{\th}{2}\Bigr)\exp(-\im\vp )\ ,\quad
\yb=R\tan\Bigl(\frac{\th}{2}\Bigr)\exp(\im\vp )\ ,
\end{equation}
where $R$ is the constant radius of $S^2$. In these coordinates
the metric on $\Sigma^{1,1}\times S^2$ has the form
\begin{eqnarray}\label{2.2}
\diff s^2&=&g_{\m\n}\diff x^\m\,\diff x^\n =
\diff s^2_{\Sigma^{1,1}}+\diff s^2_{\C P^1}
= \eta_{ij} \diff x^i\diff x^j +2\,g_{y\yb}\,\diff y\,\diff \yb
=\\
&=&(\diff x^0)^2 - (\diff x^3)^2+R^2(\diff\th^2+\sin^2\th\,\diff\vp^2)=
(\diff x^0)^2 - (\diff x^3)^2+
\frac{4R^4}{(R^2+y\yb)^2}\,\diff y\,\diff \yb\ ,\nonumber
\end{eqnarray}
Note that the volume two-form on $S^2$ reads
\begin{equation}\label{2.3}
\omega = \omega_{\th\vp}\,\diff\th\wedge\diff\vp =
R^2\,\sin\th\,\diff\th\wedge\diff\vp=
- \frac{2\im\,R^4}{(R^2+y\yb)^2}\,\diff y\wedge\diff \yb\ .
\end{equation}
Here, the bar denotes complex conjugation.

\smallskip

\noindent
{\bf Yang-Mills equations.} We consider a rank 2 Hermitian vector bundle
$\Ecal$ over $M:=\Sigma^{1,1}\times S^2$ with a gauge potential $\Acal$
on $\Ecal$ and the gauge field $\Fcal =\diff\Acal + \Acal\wedge\Acal$.
In local coordinates, $\Acal =\Acal_\m\diff x^\m$ and $\Fcal =
\sfrac{1}{2}\,\Fcal_{\m\n}\diff x^\m\wedge\diff x^\n$ with
$\Fcal_{\m\n}=\pa_\m\Acal_\n - \pa_\n\Acal_\m +[\Acal_\m , \Acal_\n ]$,
where $\pa_\m := \pa /\pa x^\m$. Both $\Acal_\m$ and $\Fcal_{\m\n}$
take values in the Lie algebra $su(2)$.

For the standard Yang-Mills action functional
\begin{equation}\label{2.4}
S=-\frac{1}{4\pi}\int_M\tr\, (\Fcal\wedge *\Fcal )=
-\frac{1}{4\pi}\int_{M}\diff^4 x\, \sqrt{g}\, \tr\,
(\Fcal_{\mu\nu}\Fcal^{\mu\nu})
\end{equation}
the equations of motion are
\begin{equation}\label{2.5}
\frac{1}{\sqrt{g}}\,\pa_\m (\sqrt{g}\,\Fcal^{\m\n}) + [\Acal_{\m}, \Fcal^{\m\n}]=0\ .
\end{equation}
Here, $*$ is the Hodge operator and $g=|\det (g_{\m\n})|$. In particular, for the
metric (\ref{2.2}) we have $\sqrt{g}=R^2\sin\th$.

\smallskip

\noindent
{\bf SO(3)-invariant gauge potential.} We want to consider the gauge fields on
the bundle $\Ecal\to\Sigma^{1,1}\times\C P^1$ which are invariant under the SU(2)
isometry group of $\C P^1 \cong\ $SU(2)/U(1). It is natural to allow for gauge
transformations to accompany the space-time SU(2) action~\cite{13}.
The $\C P^1$ dependence in this case is uniquely determined  by
the rank of the bundle $\Ecal$ and the degree $m$ of the monopole bundle
$\Lcal^{\otimes m}$ over $\C P^1$ (see e.g.~\cite{13, 15, 16}). For rank
$\Ecal=2$ and $m=1$ the gauge potential has the form
\begin{equation}\label{2.6}
\Acal = \bigl(\sfrac{1}{2}\, A +a\bigr)\, \s_3 +
\sfrac{1}{2}\,\phi\,\bar\b\,\s_+ -
\sfrac{1}{2}\, \bar\phi\,\b\,\s_-
\quad\mbox{with}\quad
\s_+{=}\begin{pmatrix}0&1\\0&0\end{pmatrix}=\s_-^\+\
\ \mbox{and}\
\s_3{=}\begin{pmatrix}1&0\\0&-1\end{pmatrix},
\end{equation}
where $A$ is an Abelian gauge potential on the
complex line bundle $\tilde L$ over $\Sigma^{1,1}$,
$\phi\in H^0(\Sigma^{1,1} , \tilde L)$
is a complex scalar, $a$ is a one-monopole gauge potential on the
bundle $\Lcal\to\C P^1$ given (locally) by
\begin{equation}\label{2.7}
a= \frac{1}{2(R^2+y\yb)}(\yb\,\diff y - y\,\diff\yb )
\end{equation}
and
\begin{equation}\label{2.8}
\b =\frac{{\sqrt 2}\,R^2}{R^2+y\yb}\, \diff y\ .
\end{equation}
Note that fields $A$ and $\phi$ depend only on coordinates on $\Sigma^{1,1}$.

\smallskip

\noindent
{\bf Symmetric gauge fields.} In local complex coordinates on
$\Sigma^{1,1}\times S^2$ the curvature $\Fcal =\diff\Acal + \Acal\wedge\Acal$
for $\Acal$ of the form (\ref{2.6}) has the field strength components
\begin{subequations}\label{2.9}
\begin{eqnarray}
\Fcal_{ij}&=& \sfrac{1}{2}\,F_{ij}\,\s_3\ ,\quad
\Fcal_{y\yb}=- \frac{1}{4}g_{y\yb} \Bigl (\frac{2}{R^2}-\phi\,\bar\phi\Bigr )\,\s_3\ ,\\
\Fcal_{i\yb}&=&\sfrac12\, {\rho}\,(D_i\phi )\,\s_+ \quad\mbox{and}\quad
\Fcal_{iy}=-\sfrac12\,{\rho}\,\overline{(D_i\phi )}\,\s_- \ ,
\end{eqnarray}
\end{subequations}
where
\begin{equation}\label{2.10}
D_i\p :=\pa_i\p + A_i\p\ , \quad \overline{D_i\p}:=\pa_i\bar\p -A_i\bar\p
\quad\mbox{and}\quad \rho :=(g_{y\yb})^{1/2}\ ,
\end{equation}
\begin{equation}\label{2.11}
F_{ij} :=\pa_iA_j - \pa_jA_i=:-\im f_{ij}
\quad\mbox{and}\quad A_j:=-\im a_j \quad \mbox{with}\quad a_j\in\R\ .
\end{equation}

\section{Abelian Higgs model on $\Sigma^{1,1}$, kinks and sphalerons}

\noindent
{\bf Abelian Higgs model.} Substituting (\ref{2.9}) into (\ref{2.4})
and performing the integral over $\C P^1$, we arrive at the action
\begin{equation}\label{3.1}
S=-\frac{1}{4\pi}\int_M\tr\, (\Fcal\wedge *\Fcal )=
{R^2}\int_{\Sigma^{1,1}}\diff^2x\,\Bigl\{\frac{1}{2}\,\bar F_{ij}F^{ij} +
\overline{D_i\p}\, D^i\p +\frac{1}{4} \Bigl (\frac{2}{R^2} - \phi\bar\phi\Bigr )^2\Bigr\}\ ,
\end{equation}
which coincides with the action functional of the Abelian Higgs model at
critical coupling $\la =1$ (see e.g.~\cite{2, 4}). In other words, there is
a one-to-one correspondence between gauge equivalence classes of solutions
($A,\p$) to the Abelian Higgs model (\ref{3.1}) and gauge equivalence classes
of SO(3)-invariant solutions $\Acal$ of the Yang-Mills equations on
$\Sigma^{1,1}\times S^2$.

\smallskip

\noindent
{\it Remark.} In the Euclidean regime, the second integral in (\ref{3.1})
is the Ginzburg-Landau free energy of a superconductor. Recall that in
the general case of this functional the last term under the integral in
(\ref{3.1}) has the form
\begin{equation}\label{3.2}
V(\p , \bar\p )= \frac{\la}{4} \Bigl (\frac{2}{R^2} - \p\bar\p\Bigr )^2
\end{equation}
and the critical value $\la =1$ of the coupling constant separates Type I
($0<\la <1$) and Type II ($\la >1$) superconductivity. The normal
conducting phase (Coulomb phase) corresponds to the limit $R\to\infty$.
Note that if instead of the spherically symmetric ansatz (\ref{2.6})
one chooses more general ansatz~\cite{17} of the form
\begin{equation}\label{3.3}
\Acal =\frac{1}{2} (\im AQ - Q\diff Q) + \frac{\im\p R}{2\sqrt{2}}(1_2+
\im Q)\diff Q - \frac{\im\bar\p R}{2\sqrt{2}}(1_2 -\im Q)\diff Q\ ,
\end{equation}
where
\begin{equation}\label{3.4}
Q=\im(\sin\th\cos (m\vp)\s_1 +\sin\th\sin (m\vp)\s_2 + \cos\th\, \s_3)
\quad\mbox{with}\quad m\in\Z\ ,
\end{equation}
then in (\ref{3.1}) one obtains the potential term (\ref{3.2}) with
\begin{equation}\label{3.5}
\la =\frac{2m^2}{m^2+1}
\end{equation}
and $\la >1$ for $m > 1$. For $m=1$ we have $\la =1$ and the potentials $\Acal$
from (\ref{2.6}) and (\ref{3.3}) are related by a gauge transformation. Note
that the Abelian Higgs model (\ref{3.1}) and its non-Abelian generalization are
integrable on compact Riemann surfaces of genus $g>1$~\cite{18}.

\smallskip

\noindent
{\bf Equations of motion.} The field equations for ($A, \p$) can be obtained
either by variation of (\ref{3.1}) or by substitution of (\ref{2.6})-(\ref{2.11})
into the Yang-Mills equations (\ref{2.5}). In both ways we obtain the
equations
\begin{equation}\label{3.6}
\pa_iF^{ij}=\frac{1}{2}\bigl(\bar\p D^j\p - \p\overline{D^j\p}\bigr)\ ,
\end{equation}
\begin{equation}\label{3.7}
D_iD^i\p +\frac{1}{2} \Bigl (\frac{2}{R^2} - \p\bar\p \Bigr )\p =0\ .
\end{equation}
These equations have interesting sphaleron solutions~\cite{19, 20} (i.e.
saddle-point static finite energy solutions~\cite{21} of the field
equations). For their study the authors of~\cite{19, 20} consider numerical
solving of classical equations of motion for the starting configuration
\begin{equation}\label{3.8}
\phi =\phi_{kink}\exp (\im\a (x^3))\quad\mbox{and}\quad A_3=-\im\pa_3\a(x^3)\ ,
\end{equation}
where $\phi_{kink}$ will be discussed in a moment and
$\a (+\infty)-\a(-\infty)=\pi$. For more details see~\cite{19, 20}.
Here, we want to concentrate on explicit solutions.

For finding explicit solutions we restrict $\phi$ to be a real scalar
field, $\bar\phi =\phi$. Then the choice $A_i=0$ solves (\ref{3.6}) and
reduces (\ref{3.7}) to the standard $\phi^4$ kink model equation
\begin{equation}\label{3.9}
\pa_i\pa^i\phi + \frac{1}{R^2}\phi - \frac{1}{2}\phi^3 =0\ ,
\end{equation}
which is the Euler-Lagrange equation for the consistently reduced action
(\ref{3.1}),
\begin{equation}\label{3.10}
S_{red}=R^2\int_{\Sigma^{1,1}}\diff^2x\left\{\pa_i\phi\,\pa^i\phi
+ \frac{1}{4}\Bigl(\frac{2}{R^2} -\phi^2\Bigr)^2\right\}\ .
\end{equation}
The energy of static configurations is
\begin{eqnarray}\label{3.11}
&E=R^2\int\limits_{b_1}^{b_2}\diff x^3\left\{(\pa_3\phi )^2
+ \frac{1}{4}\Bigl (\frac{2}{R^2} -\phi^2\Bigr)^2\right\}=\nonumber\\
&= R^2 \int\limits_{b_1}^{b_2}\diff x^3\left(  \pa_3\phi \mp
\frac{1}{2}\Bigl (\frac{2}{R^2} -\phi^2\Bigr)\right)^2
\pm
  R^2 \int\limits_{b_1}^{b_2}\diff\phi \Bigl (\frac{2}{R^2} -
  \phi^2\Bigr)
\ge \frac{8\sqrt{2}}{3R}\,|q| \ ,
\end{eqnarray}
where
\begin{equation}
q=\frac{R}{2\sqrt{2}}\int\limits_{b_1}^{b_2}\diff x^3\,\pa_3\phi
=\frac{R}{2\sqrt{2}}\bigl(\phi(b_2) - \phi(b_1)\bigr) \in \{1, 0, -1\}
\end{equation}
is the topological charge. Here, $b_1=-\infty , b_2=+\infty$ for
$\Sigma^{1,1}= \R^{1,1}$, $b_1=0, b_2=L<\infty$ for $\Sigma^{1,1}=
\R\times S^{1}$ and $\phi(b_1) = \pm\frac{\sqrt{2}}{R}, \phi(b_2) =
\pm\frac{\sqrt{2}}{R}$ are the vacua of the model (\ref{3.10}).
Note that (\ref{3.11}) is the reduction of the Yang-Mills energy
functional for static configurations. The equality in (\ref{3.11})
is attained on the BPS equations
\begin{equation}\label{BPS}
\pa_3\phi = \pm\frac{1}{2} \Bigl(\frac{2}{R^2}-\phi^2\Bigr )\ ,
\end{equation}
where the choice of the $\pm$ sign corresponds to the sign of $q$ for $q\ne 0$.

\smallskip

\noindent
{\bf Kinks.} Static solution of the BPS equation (\ref{BPS}) with the
$+$ sign is known as $\phi^4$ kink,
\begin{equation}\label{3.12}
\phi_{kink}:=\phi (x^3)=\frac{\sqrt{2}}{R}
\tanh\Bigl (\frac{1}{\sqrt{2}\,R}x^3\Bigr )\ ,
\end{equation}
which is a solution interpolating between the vacua  $-\frac{\sqrt{2}}{R}$ and
$+\frac{\sqrt{2}}{R}$ and having the topological charge $q=1$. For the antikink
with $q=-1$, interpolating between $+\frac{\sqrt{2}}{R}$ and $-\frac{\sqrt{2}}{R}$,
we have
\begin{equation}\label{3.13}
\phi_{antikink}=-\frac{\sqrt{2}}{R}
\tanh\Bigl (\frac{1}{\sqrt{2}\,R}x^3\Bigr )\ .
\end{equation}
The energy density is maximal at the point $x^3=0$ for both the kink and the
antikink. This point is interpreted as their position which can be shifted by
$x^3$-translations. The energy of the static kink is
\begin{equation}\label{3.14}
E[\phi ]=\frac{8}{3}\,\frac{\sqrt{2}}{R}\ ,
\end{equation}
and the same for the antikink.

By a Lorentz rotation of $\R^{1,1}$ one can obtain a kink
\begin{equation}\label{3.15}
\phi (x^0, x^3)=\frac{\sqrt{2}}{R}
\tanh\Bigl (\frac{\gamma}{\sqrt{2}\,R}(x^3-vx^0)\Bigr )
\end{equation}
moving with the velocity $-1<v<1$ and $\gamma :=(1-v^2)^{-1/2}$ is the
Lorentz factor~\cite{4}. For the moving kink (\ref{3.15}) the energy (\ref{3.14})
is multiplied by $\g$.

\smallskip

\noindent
{\bf Sphalerons.} For obtaining sphalerons in the model (\ref{3.9})-(\ref{3.11})
one should consider $\Sigma^{1,1}=\R\times S^1$ and static solutions of (\ref{3.9})
defined on $S^1$ with a circumference $L$ such that
\begin{equation}\label{3.16}
\phi (x^3+L)=\phi (x^3)\ .
\end{equation}
Then sphalerons are given by~\cite{14}
\begin{equation}\label{3.17}
\phi_n (x^3; k)=2k\,b(k)\,\sn[b(k)x^3; k]\quad \mbox{with}\quad
b(k)=\frac{1}{R(1+k^2)^{1/2}}\quad \mbox{and}\quad 0\le k\le 1\ .
\end{equation}
Here $\sn[x;k]$ is the Jacobi elliptic function with the period $4{\cal K}(k)$.
The periodicity condition (\ref{3.16}) is satisfied if
\begin{equation}\label{3.18}
L=\frac{4{\cal K}(k)\,n}{b(k)}
\end{equation}
for some integer $n$. Solutions (\ref{3.17}) exist if $L\ge L_n:=2\pi Rn$~\cite{14}.

By virtue of the periodic boundary condition (\ref{3.16}), the topological charge
of the sphaleron (\ref{3.17}) is zero. In fact, the configuration (\ref{3.17})
 is interpreted as a chain of $n$ kinks and $n$
antikinks alternatively and equally spaced around the circle $S^1$~\cite{4, 14}.
Energy (\ref{3.11}) of the sphaleron (\ref{3.17}) is
\begin{equation}\label{3.19}
E[\phi_n]=\frac{4n}{3R}\left [8(1+k^2)\Ecal(k)-(1-k^2)(5+3k^2){\cal K}(k)\right ]\ ,
\end{equation}
where $\Ecal(k)$ is the complete elliptic integral of the second kind~\cite{14}.
In the limit $L\to\infty$ the energy (\ref{3.19}) becomes $2n\,E[\phi ]$,
where $E[\phi ]$ is the energy (\ref{3.14}) of the single kink (\ref{3.12}).

\section{Explicit SU(2) multi-monopole configurations}

\noindent
{\bf Non-Abelian monopoles.} To obtain explicit smooth SU(2) monopole configurations
in pure Yang-Mills theory we should substitute the kink, antikink and kink-antikink
chain solutions (\ref{3.12}), (\ref{3.13}) and (\ref{3.17}) into (\ref{2.6}) and (\ref{2.9}).
In particular, the monopole gauge potential $\Acal$ reads
\begin{equation}\label{4.1}
\Acal = a\s_3 + \sfrac12\phi (\bar\b\s_+ - \b\s_-)\ ,
\end{equation}
where $a$ and $\b$ are given in (\ref{2.7}), (\ref{2.8}). We see that
$\Acal_0=0=\Acal_3$ and $\Acal_y, \Acal_{\yb}$ can be easily extracted from (\ref{4.1}).

{}For the two-form $\Fcal$ of the one-monopole gauge field we have
\begin{eqnarray}
\Fcal&=& \sfrac12\Fcal_{ij}\diff x^i\wedge\diff x^j +
\Fcal_{i\yb}\diff x^i\wedge\diff \yb +
\Fcal_{iy}\diff x^i\wedge\diff y +\Fcal_{y\yb}\diff y\wedge\diff \yb=\nonumber\\
&=& - \frac{1}{4} \Bigl(\frac{2}{R^2}-\phi^2\Bigr)\left\{ \frac{\s_1}{2\im}\,\rho\,
\diff x^2\wedge\diff x^3+\frac{\s_2}{2\im}\,\rho\,
\diff x^3\wedge\diff x^1+\frac{\s_3}{2\im}\,\rho^2\,
\diff x^1\wedge\diff x^2\right\}=\nonumber\\
\label{4.2}
&=&-\frac{1}{8} \Bigl(\frac{2}{R^2}-\phi^2\Bigr)\, \frac{\s_a}{2\im}\,
\eps^a_{bc}\,\b^b\wedge\b^c\ ,
\end{eqnarray}
where
\begin{equation}\label{4.3}
\rho =\frac{4\sqrt{2}\, R^2}{4R^2+(x^1)^2+(x^2)^2}\ ,\quad
\b^1:=\rho\,\diff x^1\ ,\quad \b^2:=\rho\,\diff x^2\quad
\mbox{and}\quad\b^3:=\diff x^3\ .
\end{equation}
Here, $\{\b^a\}$ forms the nonholonomic basis of one-forms on
$\T\times S^2$ with $\T=\R$ or $\T=S^1$. In derivation of (\ref{4.2})
we used the fact that the kink solutions (\ref{3.12}) satisfy the BPS equation
(\ref{BPS}) with the $+$ sign. Thus, $\Fcal$
from (\ref{4.2}) is a smooth $su(2)$ gauge field on $\R\times S^2$ with finite
energy given by (\ref{3.14}) for $\phi$ from (\ref{3.12}).
Similarly, substituting (\ref{3.13}) into (\ref{2.9}), we obtain antimonopole
gauge field which coincides in form with (\ref{4.2}) for $x^3\to -x^3$.
\smallskip

\noindent
{\bf Topological charges.} To understand the topological nature
of the solution (\ref{4.2}), we consider the asymptotic behaviour of the
gauge potential (\ref{4.1}) which follows from the explicit form (\ref{3.12})
of $\phi$. Note that for the Euclidean space $\R\times S^2$ `infinity' is the
disconnected space
\begin{equation}\label{4.4}
S^2\times Z_2 = S^2_- \cup S^2_+ \ ,
\end{equation}
where $S^2_{\pm} = S^2\times \{x^3 = \pm \infty\}$ are
two-spheres at $x^3 = \pm \infty$. Introducing notation
$\Acal_{\pm}:= \Acal (x^3 = \pm \infty)$, from (\ref{4.1})
we obtain
\begin{equation}\label{t4.5}
\Acal_- = h\,\diff h^{-1}
\quad\mbox{and}\quad
\Acal_+ = h^{-1}\diff h
\quad\Rightarrow\quad
\Fcal_{\pm}=0\ (\mbox{vacuum})\ ,
\end{equation}
where
\begin{equation}\label{t4.6}
h = \frac{1}{\sqrt{R^2+y\bar y}}\begin{pmatrix} R&\bar y\\-y&R\end{pmatrix} \in \mbox{SU}(2)\ .
\end{equation}

For further clarification, we use $h$ for the gauge
transformation of $\Acal_-$ to zero, i.e. consider
\begin{equation}\label{t4.7}
\tilde\Acal := h^{-1}\Acal\, h + h^{-1}\diff h
\quad\Rightarrow\quad
\tilde\Acal_- =0
\quad\mbox{and}\quad
\tilde\Acal_+ = g^{-1}\diff g\ ,
\end{equation}
where
\begin{equation}\label{t4.8}
g:= h^2 = \frac{1}{(R^2+y\bar y)}\begin{pmatrix} R^2-y\bar y& 2R\bar y\\
-2R y& R^2-y\bar y\end{pmatrix} =
\begin{pmatrix}\cos\theta&\sin\theta\exp(i\varphi)\\
-\sin\theta\exp(-i\varphi)& \cos\theta \end{pmatrix} \in \mbox{SU}(2)\ .
\end{equation}
In (\ref{t4.8}) we recognize the map $g: S^2_+ \to S^2 \subset$ SU(2)
of degree 1 in full agreement with the fact that the topological charge of
the kink $\phi$ is $q=1$. Thus, if we consider $x^3$ as Euclidean time of Yang-Mills
theory in $2+1$ dimensions, then our solution (\ref{4.1}),(\ref{4.2}) is
interpreted as the one-instanton configuration describing transition between
the vacua $\tilde\Acal_- =0$ and $\tilde\Acal_+ = g^{-1}\diff g$. The (finite)
action for this `instanton' configuration is given by formula (\ref{3.14}).
On the other hand, from the viewpoint of Yang-Mills theory in $3+1$ dimensions,
the solution (\ref{4.1}), (\ref{4.2}) describes a finite-energy static one-monopole
configuration with the magnetic field strength (\ref{4.2}). In fact, this is
the standard equivalence of instanton and monopole interpretations of
finite-action Yang-Mills (-Higgs) topological solitons in $3+0$ dimensions.
It is especially obvious in the case of Yang-Mills theory on the space $\R\times S^2$.

For antimonopole, related with antikink, one can repeat the above calculations
and obtain after the gauge transformation ${\Acal}\mapsto {\Acal}^\prime:=
h\,\Acal\, h^{-1} + h\,\diff h^{-1}$ that
\begin{equation}\label{t4.9}
{\Acal}^\prime_- = 0
\quad\mbox{and}\quad
{\Acal}^\prime_+ = g\,\diff g^{-1} =g\,\diff g^{\+}.
\end{equation}
Hence, for the antimonopole we have the map $g^{\+} : S^2_+ \to S^2 \subset$ SU(2)
of degree $q = -1$, as expected. Note again that the topological charges of
monopole and antimonopole on the space $\R\times S^2$ coincide with the topological
charges of kink and antikink, respectively.

\smallskip

\noindent
{\bf Dyons.} The moving kink (\ref{3.15}) corresponds to a non-Abelian dyon.
It has the same form (\ref{4.1}) of the gauge potential but for $\Fcal$ we
obtain
\begin{equation}\label{4.5}
\Fcal{=}\sfrac{1}{8}\bigl(\sfrac{2}{R^2}-\phi^2\bigr )
\bigl\{\g v \im\s_1\r \diff x^0\wedge\diff x^2 -
\g v \im\s_2\r\diff x^0\wedge\diff x^1+
\g \im\s_1\r\diff x^2\wedge\diff x^3+ \g \im\s_2\r \diff x^3\wedge\diff x^1
+ \im\s_3\r^2\diff x^1\wedge\diff x^2\bigr\}\ .
\end{equation}
Thus, for the dyon solution (\ref{4.5}) the electric components $\Fcal_{01}$ and
$\Fcal_{02}$ are nonvanishing if $v\ne 0$. Its energy is the same as
of moving kink.

\smallskip

\noindent
{\bf Monopole-antimonopole chains.}
For the solution (\ref{3.17}) the equations (\ref{BPS}) are not valid and we have
\begin{equation}\label{4.6}
\Fcal =\sfrac{1}{4}(\pa_3\p )\, \im\s_1\,\b^2\wedge\b^3 +
\sfrac{1}{4}(\pa_3\p )\, \im\s_2\,\b^3\wedge\b^1 +
\sfrac{1}{8} \bigl(\sfrac{2}{R^2}-\phi^2\bigr )\,\im\s_3\,\b^1\wedge\b^2
\end{equation}
with $\b^a$ given in (\ref{4.3}). This field describes a chain of $n$ monopole-antimonopole
pairs on $S^1\times S^2$ with the energy (\ref{3.19}).

\section{Explicit monopole solutions in SU($N$) gauge theory}

\noindent
Here we generalize the results of the previous Sections to the case of SU($N$)
Yang-Mills theory. So, we consider now a rank $N$ Hermitian vector bundle $\Ecal$
over $M:=\Sigma^{1,1}\times S^2$ with fields $\Acal_\mu$ and $\Fcal_{\m\n}$ in (\ref{2.4}),
(\ref{2.5}) taking values in the Lie algebra $su(N)$.

\smallskip

\noindent
{\bf SO(3)-invariance.} For imposing SO(3)-invariance on our $\Acal$ and $\Fcal$, we
choose a particular case from those ones considered in~\cite{16} with $k_0=...=k_m=1$
and therefore
$N=m+1$, where $m$ is the maximal first Chern number of the Dirac monopole bundles
$\Lcal^{m-2\ell}:=\Lcal^{\otimes(m-2\ell )}\to\C P^1$ with $0\le\ell\le m$.
On each such line bundle $\Lcal^{m-2\ell}$ we have the unitary connection
\begin{equation}\label{5.1}
a^{m-2\ell } = (m-2\ell )a\ ,
\end{equation}
where $a$ is given in (\ref{2.7}), $\ell =0,...,m$ and $m:=N-1$. The simplest choice
generalizing (\ref{2.6}) is
\begin{equation}\label{5.2}
\Acal = \sfrac12\, A^{(m)} + a^{(m)} + \sfrac12\, \Phi_m\bar\b - \sfrac12\, \Phi_m^\+\b\ ,
\end{equation}
where
\begin{equation}\label{5.3}
A^{(m)}=\mbox{diag} (A^0,..., A^\ell ,..., A^m)\quad\mbox{and}\quad
a^{(m)}=\mbox{diag} (a^m,..., a^{m-2\ell } ,..., a^{-m})
\end{equation}
are diagonal traceless (i.e. $\sum\limits^m_{\ell =0}A^\ell =0$) $N\times N$ matrices
and
\begin{equation}\label{5.4}
\Phi_m=\begin{pmatrix}0&\phi_1&...&0\\\vdots&0&\ddots&\vdots\\
\vdots&&\ddots&\phi_m\\0&...&...&0\end{pmatrix}
\end{equation}
For more general ans\"atze see~\cite{16}.

For $\Acal$ of the form (\ref{5.2}) the field strength components are
\begin{subequations}\label{5.5}
\begin{eqnarray}
\Fcal_{ij}&=& \sfrac{1}{2}\,F_{ij}^{(m)}=\sfrac{1}{2}(\pa_i A_j^{(m)}- \pa_j A_i^{(m)})\ ,\quad
\Fcal_{y\yb}=- \frac{1}{4}g_{y\yb} \bigl (\frac{2}{R^2}\Ups_m - [\Phi_m,\Phi^\+_m]\bigr )\ ,\\
\Fcal_{i\yb}&=&\sfrac12\, {\rho}\,(D_i\Phi_m )\quad\mbox{and}\quad
\Fcal_{iy}=-\sfrac12\,{\rho}\,(D_i\Phi_m )^\+ \ ,
\end{eqnarray}
\end{subequations}
where
\begin{equation}\label{5.5c}
D_i\Phi_m:=\pa_i\Phi_m + [A^{(m)}, \Phi_m]\ , \quad \Ups_m=\mbox{diag} (m,...,m-2\ell ,..., -m)
\end{equation}
and $\rho$ is given in (\ref{2.2}), (\ref{2.10}).

\smallskip

\noindent
{\bf Matrix $\Phi^4$ kink model.} Substituting (\ref{5.2})-(\ref{5.5}) into (\ref{2.4}) and
integrating over $\C P^1$, we obtain the non-Abelian Higgs model
\begin{eqnarray}
S&=&-\frac{1}{4\pi}{\int_M}\tr(\Fcal{\wedge}*\Fcal )=\nonumber\\\label{5.6}
&=&R^2{\int_{\Sigma^{1,1}}}\diff^2x\,\tr\Bigl\{\frac{1}{4}(F_{ij}^{(m)})^\+ F^{(m)ij}+
(D_i\Phi_m)^\+ D^i\Phi_m+\frac{1}{8}\bigl (\frac{2}{R^2}\Ups_m -[\Phi_m,\Phi_m^\+]\bigr )^2\Bigr\}\ ,
\end{eqnarray}
which is equivalent to a model of $m$ interacting complex scalar fields and $m$ Abelian gauge fields.
Similarly to the SU(2) case, we can consistently impose the reality condition
$\bar{\phi_\ell}=\phi_\ell$ for $\ell=1,...,m$ and then put $A^{(m)}=0$. This choice reduces
(\ref{5.6}) to the action functional of the matrix $\Phi^4$ kink model
\begin{equation}\label{5.7}
S_{red}=R^2\int_{\Sigma^{1,1}}\diff^2x\, \tr\left\{\pa_i\Phi_m\,\pa^i\Phi_m^\+
+ \frac{1}{8}\bigl(\frac{2}{R^2}\Ups_m -[\Phi_m,\Phi_m^\+]\bigr)^2\right\}\ ,
\end{equation}
describing interacting $\phi^4$-type kinks.

{}From (\ref{5.7}) we obtain the matrix field equation
\begin{equation}\label{5.8}
\pa_i\pa^i\Phi_m + \frac{1}{R^2}\Phi_m-
\frac{1}{4}\bigl [[\Phi_m,\Phi_m^\+], \Phi_m\bigr]=0\ ,
\end{equation}
which is equivalent to the linked equations
\begin{equation}\label{5.9}
\pa_i\pa^i\p_\ell + \frac{1}{R^2}\p_\ell + \frac{1}{4}\bigl(\p^2_{\ell -1}-2\p^2_\ell +
\p^2_{\ell +1}\bigr) \p_\ell=0
\end{equation}
with $\ell =1,...,m$ and $\p_0:=0=:\p_{m+1}$.

\smallskip

\noindent
{\bf Explicit kink-type solutions.} From (\ref{5.7}) and (\ref{5.8}) one can see that the
vacua are given by
\begin{equation}\label{5.10}
\frac{\sqrt{2}}{R}\Phi^0_m\quad\mbox{with}\quad\Phi^0_m\ :\quad \p^0_\ell
= \pm\sqrt{\ell (m-\ell +1)}\quad\mbox{for}\quad \ell =1,...,m
\end{equation}
and therefore we have $2^m$ distinct vacua for different combinations of signs in (\ref{5.10}).

Simplest static solution of eq. (\ref{5.8}) is
\begin{equation}\label{5.11}
\Phi_m=\Phi^0_m\frac{\sqrt{2}}{R}\tanh\Bigl(\frac{1}{\sqrt{2}R}x^3\Bigr)\ ,
\end{equation}
where $\Phi^0_m$ is anyone from $2^m$ vacua. We are not able to describe the general
solution of eqs. (\ref{5.9}) but can give many explicit examples. In particular, nontrivial
smooth solution can be obtained by taking $\p_\ell$ vanishing for $\ell$ from a subset
$I_0$ of indices $I=\{1,...,m\}$ and as a kink or an antikink (with the topological charge $q_\ell=\pm 1$) for other values of $\ell\in I\setminus I_0$. For instance, we can take
\begin{eqnarray}
m{=}2&:&\p_1{=}q_1\frac{\sqrt{2}}{R}\tanh\Bigl(\frac{1}{\sqrt{2}R}(x^3-a_1)\Bigr)
\quad\mbox{and}\quad\p_2{=}0\label{5.12}\\
m{=}3&:&\p_1{=}q_1\frac{\sqrt{2}}{R}\tanh\Bigl(\frac{1}{\sqrt{2}R}(x^3{-}a_1)\Bigr)
, \ \p_2{=}0, \
\p_3{=}q_3\frac{\sqrt{2}}{R}\tanh\Bigl(\frac{1}{\sqrt{2}R}(x^3{-}a_3)\Bigr)\label{5.13}\\
m{=}2r&:&\p_1{=}q_1\frac{\sqrt{2}}{R}\tanh\Bigl(\frac{1}{\sqrt{2}R}(x^3{-}a_1)\Bigr)
, \ \p_2{=}0, \
\p_3{=}q_3\frac{\sqrt{2}}{R}\tanh\Bigl(\frac{1}{\sqrt{2}R}(x^3{-}a_3)\Bigr)\ ,\nonumber\\
&&\quad \ldots\quad
\p_{m-1}=q_{m-1}\frac{\sqrt{2}}{R}\tanh\Bigl(\frac{1}{\sqrt{2}R}(x^3{-}a_{m-1})\Bigr), \
\p_m=0\label{5.14}\\
m{=}2r{+}1&:&\p_1{=}q_1\frac{\sqrt{2}}{R}\tanh\Bigl(\frac{1}{\sqrt{2}R}(x^3{-}a_1)\Bigr),
 \ \p_2{=}0, \
\p_3{=}q_3\frac{\sqrt{2}}{R}\tanh\Bigl(\frac{1}{\sqrt{2}R}(x^3{-}a_3)\Bigr)\ ,\nonumber\\
&&\quad \ldots\quad \p_{m-1}=0, \
\p_{m}=q_{m}\frac{\sqrt{2}}{R}\tanh\Bigl(\frac{1}{\sqrt{2}R}(x^3{-}a_{m})\Bigr) \ .
\label{5.15}
\end{eqnarray}
Here, each $\p_\ell\ne 0$ describes a kink (for $q_\ell =1$) or antikink (for $q_\ell =-1$)
located at the point $a_\ell\in\R$. In (\ref{5.14}), (\ref{5.15}), one can consider any $N=m+1>1$,
including $N>>1$. The energy of any such configuration is infinite due to the contribution of
saddle points $\phi_{\ell} = 0$. However, it can be `regularized' by subtracting a constant from
the potential energy density.

\smallskip

\noindent
{\bf Sphaleron-type solutions.} This kind of static periodic solutions appear on
$\Sigma^{1,1}=\R\times S^1$. The simplest one is
\begin{equation}\label{5.16}
\Phi_m=\Phi^0_m\p_n(x^3; k)\ ,
\end{equation}
where the sphaleron $\p_n$ is given by formulae (\ref{3.17})-(\ref{3.19}). Furthermore,
suppose we were able to choose $0\le k_\ell\le 1$ and $n_\ell$ such that $\p_\ell$
have the same period (\ref{3.18}) for different pairs $(n_\ell , k_\ell )$. Then
we can introduce more general than (\ref{5.16}) sphaleron-type solutions
\begin{subequations}\label{5.17}
\begin{eqnarray}
m{=}2r&:& \p_1{=}\p_{n_1}(x^3; k_1),\ \p_2{=}0,\ \ldots\ ,
\p_{m-1}{=}\p_{n_{m-1}}(x^3; k_{m-1}),\ \p_m=0\\
m{=}2r{+}1&:&\p_1{=}\p_{n_1}(x^3; k_1),\ \p_2{=}0, \ \ldots \ ,
\p_{m-1}=0, \
\p_{m}= \p_{n_m}(x^3; k_m)\ ,
\end{eqnarray}
\end{subequations}
whose substitution into $\Phi_m$ does not reproduce the factorized form (\ref{5.16}).

\smallskip

\noindent
{\bf Explicit SU($N$) monopole configurations.} To obtain static monopole configuration in
SU($N$) pure Yang-Mills theory on $\R\times\T\times S^2$, we should substitute into
(\ref{5.2})-(\ref{5.5}) with $A^{(m)}{=}0$ any solution $\Phi_m$ of equation (\ref{5.8})
smooth on $\T=\R$ or $\T=S^1$.

For the simplest solutions (\ref{5.11}) and (\ref{5.16}), the explicit form of $\Fcal$
can be obtained from (\ref{4.2}) or (\ref{4.6}) by substituting
\begin{equation}\label{5.18}
\s_+\to\Phi_m^0\ ,\quad \s_-\to(\Phi_m^0)^\+\quad\mbox{and}\quad
\s_3\to\Ups_m\ ,
\end{equation}
i.e. via embedding the Lie algebra $su(2)$ into $su(N)$ since $\im(\Phi_m^0+(\Phi_m^0)^\+)$,
$\Phi_m^0-(\Phi^0_m)^\+$ and $\im\Ups_m$ form the generators of the $N$-dimensional
representation of SU(2). Therefore, the monopole configurations on $\T\times S^2$ obtained
via (\ref{5.18}) are $su(2)$ solutions in disguise.

More interesting solutions, not reduced to the $su(2)$-case, will be obtained if we substitute
into (\ref{5.2}) and (\ref{5.5}) the sandwich-type solutions (\ref{5.12})-(\ref{5.15}) on $\R$
or (\ref{5.17}) on $S^1$. We will not write down the explicit form of these smooth
configurations since they are obtainable simply by the substitution of (\ref{5.12})-(\ref{5.15})
or (\ref{5.17}) into (\ref{5.5}). Their topological charges can be obtained by summation of
charges of kink/antikink `entering' e.g. in (\ref{5.14}), (\ref{5.15}) by virtue of the
kink/monopole correspondence discussed in Section 4. These solutions describe non-Abelian
magnetic flux tubes extended along the $x^3$-axis since their total energy is infinite as we
noted before.

\smallskip

\noindent
{\bf Towards supersymmetric monopoles.} It is of interest to generalize our monopole solutions
to the supersymmetric case. For this, one should consider $\Ncal$-extended supersymmetric Yang-Mills
(SYM) theory on $\Sigma^{1,1}\times S^2$ and impose an SO(3)-invariance condition on fermionic and
scalar fields from a proper supermultiplet. For instance, one can consider $\Ncal =4$ SYM model
(maximal supersymmetry) and reduce it to a fermionic matrix $\Phi^4$-type kink model in $1+1$
dimensions. Also one can consider the dimensional reduction of $\Ncal =3$ SYM theory
(equivalent to $\Ncal =4$)  which is integrable by twistor methods~\cite{22} (see~\cite{23}
for recent reviews and references). The case $\Ncal =1$ and $\Ncal =2$ can also be considered.

Recall that there exist sphaleron-type solutions not only in $\p^4$ kink model but also in the
Abelian Higgs model (\ref{3.1}) on $\Sigma^{1,1}$ with $A\ne 0$~\cite{19, 20}. Therefore, the results
of Refs.~\cite{19, 20} on creation of kink-antikink pairs, nonperturbative nonconservation of
fermion quantum numbers etc. can be uplifted to $\Ncal$-extended SYM theory on $\Sigma^{1,1}\times S^2$.
For instance, the creation of kink-antikink pairs will be equivalent to a creation of
monopole-antimonopole pairs. Similar results for the case $N>2$ are expectable.

\newpage

\section{Instantons on $\R\times S^3$ and instanton-antiinstanton chains on $S^1\times S^3$}

\noindent
{\bf Manifold $\T\times S^3$.}  Let us consider the Euclidean space $\T\times S^3$, where $\T$
is $\R$ or $S^1$. The three-dimensional sphere can be described via the embedding $S^3\subset\R^4$
by the equation
\begin{equation}\label{6.1}
\de_{\m'\n'}\, x^{\m'}x^{\n'}=R^2\ ,
\end{equation}
where $\m', \n', ... =1,...,4$. On $S^3$ one can introduce left-invariant one-forms $\{e^a\}$
as (cf.~\cite{24})
\begin{equation}\label{6.2}
e^a:=\frac{\sqrt{2}}{R}\, \bar\eta^a_{\m'\n'}\, x^{\m'}\diff x^{\n'}\ ,
\end{equation}
where $\bar\eta^a_{\m'\n'}$ are the anti-self-dual 't~Hooft tensors and $a,b,... =1,2,3$. These one-forms
satisfy the Maurer-Cartan equations
\begin{equation}\label{6.3}
\diff e^a - \frac{1}{\sqrt{2}R}\, \eps^a_{bc}\, e^b\wedge e^c =0\ .
\end{equation}
Introducing $e^4:=\diff x^4=\diff\tau$, we can write the metric on $\T\times S^3$ in the form
\begin{equation}\label{6.4}
\diff s^2 = \de_{ab}\, e^a\, e^b + e^4\, e^4\ .
\end{equation}

\smallskip

\noindent
{\bf SO(4)-invariant gauge fields.} Let us consider a rank $N$ Hermitian vector bundle
$\Ecal$ over $\T\times S^3$ with a gauge potential $\Acal$ on $\Ecal$ and the gauge field
$\Fcal =\diff\Acal + \Acal\wedge\Acal$, both with values in the Lie algebra $su(N)$.
Since $S^3=\,$SO(4)/SO(3) is a homogeneous SO(4)-space, we can introduce SO(4)-invariant
connection $\Acal$ and its curvature $\Fcal$. In the `temporal gauge' $\Acal_\tau =0$
this connection has the form (cf.~\cite{25})
\begin{equation}\label{6.5}
\Acal = \sfrac12\, X_ae^a\ ,
\end{equation}
where $X_a=X_a(\tau )$ for $a=1,2,3$ are arbitrary $su(N)$-valued functions of $\tau$.

For the SO(4)-invariant curvature $\Fcal$ of $\Acal$ we have
\begin{equation}\label{6.6}
\Fcal =\diff\Acal + \Acal\wedge\Acal= -\frac{1}{2}\, \dot X_a\, e^a\wedge\diff\tau +
\frac{1}{2}\, \Bigl (\frac{1}{\sqrt{2}R}\,\eps^c_{ab}\, X_c + \frac{1}{4}\, [X_a, X_b]\Bigr )\,
e^a\wedge e^b
\end{equation}
and therefore
\begin{equation}\label{6.7}
\Fcal_{ab}=\frac{1}{\sqrt{2}R}\,\eps_{abc}X_c + \frac{1}{4}\,[X_a, X_b]\quad\mbox{and}\quad
\Fcal_{4a}=\frac{1}{2}\,\dot X_a :=\frac{1}{2}\,\frac{\diff X_a}{\diff\tau}\ .
\end{equation}
Note that we have chosen normalization coefficients in (\ref{6.2}) and (\ref{6.5}) so
that our reduced field equations be in conformity with those from Section 3 and Section 4.

\smallskip

\noindent
{\bf Matrix equations.} On the Euclidean space $\T\times S^3$ one can consider self-dual
Yang-Mills (SDYM) and anti-self-dual Yang-Mills (ASDYM) equations,
\begin{eqnarray}
\mbox{SDYM}&:& \Fcal_{a4}=\frac{1}{2}\,\eps_{abc}\,\Fcal_{bc}\label{6.8}\\
\mbox{ASDYM}&:& \Fcal_{a4}=-\frac{1}{2}\,\eps_{abc}\,\Fcal_{bc}\label{6.9}
\end{eqnarray}
solutions of which are instantons and antiinstantons, respectively.

Substitution of (\ref{6.7}) into equations (\ref{6.8}) and (\ref{6.9}) reduce them to the
first order equations:
\begin{eqnarray}
\dot X_a &=& -\frac{\sqrt{2}}{R}\,X_a - \frac{1}{4}\, \eps_{abc}\, [X_b, X_c]
\quad\mbox{for}\quad\mbox{SDYM}\ ,\label{6.10}\\
\dot X_a &=& \frac{\sqrt{2}}{R}\,X_a + \frac{1}{4}\, \eps_{abc}\, [X_b, X_c]
\quad\mbox{for}\quad\mbox{ASDYM}\ .\label{6.11}
\end{eqnarray}
On the other hand, the full Yang-Mills equations (\ref{2.5}) reduce to the second order
equations
\begin{equation}\label{6.12}
\ddot X_a=\frac{2}{R^2}X_a+\frac{3}{2\sqrt{2}R}\, \eps_{abc}\, [X_b, X_c]+
\frac{1}{4}\,\bigl [X_b, [X_a,X_b]\bigr]\ ,
\end{equation}
which obviously follow from both equations (\ref{6.10}) and (\ref{6.11}) but not vice versa.
These equations describe a simple matrix model which is the Euclidean version of the
truncation $\Ncal =4\to\Ncal =0$ of the well-known plane wave matrix model. For its
relation with $\Ncal =4$ SYM theory on $\R\times S^3$ see e.g.~\cite{26} and references
therein. For gravity dual description see e.g.~\cite{27}.

\smallskip

\noindent
{\bf Toda-like equations.} First order equations (\ref{6.10}) and (\ref{6.11}) are integrable
due to integrability of initial SDYM and ASDYM equations (\ref{6.8}) and (\ref{6.9}). They
can be reduced further to equations generalizing Toda chain equations as we describe below.

Let $\{H_\a , E_\a , E_{-\a}\}$ be the Chevalley basis for the Lie algebra $su(N)$ with the
commutation relations
\begin{equation}\label{6.13}
[H_\a , H_\b ]=0\ ,\ \ [E_\a , E_{-\b}]=\de_{\a\b}H_\b\ ,\ \ [H_\a , E_\b ]= K_{\a\b}E_\b
\quad\mbox{and}\quad [H_\a , E_{-\b}]= - K_{\a\b}E_{-\b}\ ,
\end{equation}
where $K_{\a\b}$ are components of the Cartan matrix and $\a , \b , ...=1,...,m=N-1$.
We choose for $\{X_a\}$ in (\ref{6.10}) the (algebraic) ansatz (cf.~\cite{28})
\begin{equation}\label{6.14}
X_1=\sum\limits^m_{\a =1}\rho_\a (E_\a - E_{-\a})\ ,\quad
X_2=-\im\sum\limits^m_{\a =1}\rho_\a (E_\a + E_{-\a})
\quad\mbox{and}\quad X_3=\im\sum\limits^m_{\a =1}f_\a H_\a\ ,
\end{equation}
where $\r_\a$ and $f_\a$ are arbitrary real-valued functions of $\tau$.
It is not difficult to see that after substituting (\ref{6.14}) into (\ref{6.10}) and using
(\ref{6.13})  we obtain the equations
\begin{equation}\label{6.15}
\ddot\phi_\a +\frac{\sqrt{2}}{R}\,\dot\phi_\a=\sum\limits^m_{\b =1} K_{\a\b}\,\exp\phi_\b -
\frac{4}{R^2}\ ,
\end{equation}
where
\begin{equation}\label{6.16}
\p_\a := 2\log\rho_\a\quad\mbox{and}\quad f_\a=\sum\limits_{\b =1}^{m} K_{\a\b}^{-1}\Bigl(\dot\phi_\b
+\frac{2\sqrt{2}}{R}\Bigr)\quad\mbox{with}\quad\sum\limits^m_{\g =1} K_{\a\g}^{-1}K_{\g\b}=\de_{\a\b}\ .
\end{equation}
The standard $A_m$ Toda chain equations follow from (\ref{6.15}) in the limit $R\to\infty$.
Note that in (\ref{6.15}) one can consider the limit $N=m+1\to\infty$ and obtain Toda-like lattice
equations.

\smallskip

\noindent
{\bf Some explicit solutions.} Although (\ref{6.15}) are integrable equations, we will not try
to construct their general solutions here. Instead, we restrict ourselves to some simple solutions of eqs.
(\ref{6.12}) and (\ref{6.15}) related with the $\phi^4$ kink equation. Namely, we consider the ansatz
\begin{equation}\label{6.17}
X_a = (\phi - c) T_a\ ,
\end{equation}
where $\phi =\phi(\tau )$ is a function of $\tau$, $c$ is a constant and $T_a$'s are generators
of $N$-dimensional representation of SU(2). Substituting (\ref{6.17}) into (\ref{6.12}), we obtain
the equation
\begin{equation}\label{6.18}
\ddot\phi +\Bigl(\frac{{2}c}{R^2}-\frac{3c^2}{\sqrt{2}R}+\frac{c^3}{2}\Bigr )-
\Bigl(\frac{2}{R^2}-\frac{6c}{\sqrt{2}R}+\frac{3c^2}{2}\Bigr )\phi -
\Bigl(\frac{3}{\sqrt{2}R}-\frac{3c}{2}\Bigr )\phi^2 - \frac{1}{2}\phi^3=0\ ,
\end{equation}
which for $c=\sqrt{2}/R$ reduces to the equation
\begin{equation}\label{6.19}
\ddot\phi + \frac{1}{R^2}\phi - \frac{1}{2}\phi^3=0\ ,
\end{equation}
coinciding with the static form of $\phi^4$-kink equation (\ref{3.9}) after substituting
$x^3\to\tau$. Thus, solutions (\ref{3.12}) and (\ref{3.13}) with $x^3\to\tau$ produce
respectively the instanton and the antiinstanton (via (\ref{6.5}) and (\ref{6.17})) solutions
on $\R\times S^3$, while (\ref{3.17}) gives a chain of $n$ instantons and $n$ antiinstantons
on the space $S^1\times S^3$. Recall that the circumference of $S^1$ is $L$ and the
radius of $S^3$ is $R$.

In particular, for the one-instanton solution on $\R\times S^3$ we obtain
\begin{equation}\label{6.20}
\Acal = \frac{1}{2}\Bigl (\phi - \frac{\sqrt{2}}{R}\Bigr )\,T_ae^a
\quad\mbox{and}\quad\Fcal = \frac{1}{2}\Bigl (\frac{2}{R^2}-\phi^2\Bigr )
\Bigl (\diff\tau\wedge  e^a - \frac{1}{4}\eps^a_{bc}e^b\wedge e^c \Bigr )  T_a\ ,
\end{equation}
where
\begin{equation}\label{6.21}
\phi = \frac{\sqrt{2}}{R}\tanh \Bigl (\frac{\tau}{\sqrt{2}R}\Bigr )\ .
\end{equation}
Note that (\ref{6.17}) corresponds to the ansatz
\begin{equation}\label{6.22}
\r_\a^2=\exp\p_\a = \frac{1}{2}\a (m-\a +1) \Bigl (\p -\frac{\sqrt{2}}{R}\Bigr )
\quad\mbox{and}\quad
f_\a=-\a  (m-\a +1) \Bigl (\p -\frac{\sqrt{2}}{R}\Bigr )
\end{equation}
which reduces (\ref{6.10}), (\ref{6.14}) and (\ref{6.15}) to the equation (\ref{BPS}) with
$x^3\to \tau$, and (\ref{6.21}) is a solution of this first order equation.

Similarly, substituting into (\ref{6.5})-(\ref{6.7}) and (\ref{6.17}) the solution
\begin{equation}\label{6.23}
\p_n(\tau ; k)=2k\,b(k)\,\sn [b(k)\tau ;\, k]
\end{equation}
of eq. (\ref{6.19}) on $S^1$ with all parameters given in (\ref{3.17})-(\ref{3.19}),
we obtain a finite-action configuration which is interpreted as a chain of $n$
instanton-antiinstanton pairs sitting on $S^1_L\times S^3_R$. Note that applications
of such kind configurations on the Euclidean space $\R^4$ were considered e.g. in
\cite{9, 10}. It would be interesting to construct more general solutions of eqs.
(\ref{6.12}) and (\ref{6.15}). Another possible generalization is to consider noncommutative
multi-instantons (see e.g.~\cite{29} for 't Hooft type instantons on the noncommutative
$\R^4$) on a proper deformation of the space $\T\times S^3$. Note that this space is a
Lie group both for $\T=\R$ and $\T =S^1$. Therefore, one can consider not only various
deformations of spheres (see e.g.~\cite{30}) but also a quantum group type deformation
of the space $\T\times S^3$.

\end{document}